\documentclass[sigconf]{acmart}
\renewcommand\footnotetextcopyrightpermission[1]{}

\acmConference[SOSP '25]{Symposium on Operating Systems Principles}{October 13--16, 2025}{Seoul, Republic of Korea}
\acmYear{2025}
\copyrightyear{2025}

\AtBeginDocument{%
  }
\AtBeginDocument{%
  }

\usepackage{url}
\usepackage{subcaption}
\usepackage{listings}
\usepackage{xcolor}
\usepackage{hyperref}

\lstdefinelanguage{PythonStyle}{
    language=Python,
    basicstyle=\ttfamily\small,
    keywordstyle=\color{blue},
    stringstyle=\color{orange},
    commentstyle=\color{gray}\itshape,
    morekeywords={RAGraph, Server, START, END}
}

\newcommand{\tool}[0]{{HedraRAG}}
\newcommand{\graph}[0]{{RAGraph}}

\begin{document}

\title{{\tool}: Coordinating LLM Generation and Database Retrieval in Heterogeneous RAG Serving}

\author{Zhengding Hu}
\affiliation{
  \institution{University of California San Diego}
  \city{}
  \country{}
}

\author{Vibha Murthy}
\affiliation{
  \institution{University of California San Diego}
  \city{}
  \country{}
}

\author{Zaifeng Pan}
\affiliation{
  \institution{University of California San Diego}
  \city{}
  \country{}
}

\author{Wanlu Li}
\affiliation{
  \institution{University of California San Diego}
  \city{}
  \country{}
}

\author{Xiaoyi Fang}
\affiliation{
  \institution{RegAilator Inc}
  \city{}
  \country{}
}

\author{Yufei Ding}
\affiliation{
  \institution{University of California San Diego}
  \city{}
  \country{}
}

\author{Yuke Wang}
\affiliation{
  \institution{Rice University}
  \city{}
  \country{}
}



\begin{abstract}

This paper addresses emerging system-level challenges in heterogeneous retrieval-augmented generation (RAG) serving, where complex multi-stage workflows and diverse request patterns complicate efficient execution. We present {\tool}, a runtime system built on a graph-based abstraction that exposes optimization opportunities across stage-level parallelism, intra-request similarity, and inter-request skewness. These opportunities are realized through dynamic graph transformations, such as node splitting, reordering, edge addition, and dependency rewiring, applied to wavefronts of subgraphs spanning concurrent requests. The resulting execution plans are mapped onto hybrid CPU–GPU pipelines to improve resource utilization and reduce latency. Evaluations across a wide range of RAG workflows demonstrate speedups exceeding 1.5× and reaching up to 5× over existing frameworks, showcasing the effectiveness of coordinated generation and retrieval in serving environments.

\end{abstract}


\maketitle

\section{Introduction}

Large Language Models (LLMs) are fundamentally transforming the landscape of the AI landscape. With their rapidly expanding capabilities, LLMs have been widely adopted in knowledge-intensive scenarios \cite{kandpal2023strugglelongknowledge, cao2024ragaccurate}, including everyday question answering \cite{kamalloo2023everydayqa}, engineering applications \cite{song2023llmplan}, and scientific research \cite{zhang2024honeycomb}. This has created a growing demand to integrate LLMs with external knowledge sources, containing information beyond the models' original training cut-off and built-in knowledge.

In response to the growing need for external knowledge integration, Retrieval-Augmented Generation (RAG) \cite{lewis2020rag} has emerged as a promising solution. It enables LLMs to access external knowledge without the prohibitive cost of pretraining \cite{kaplan2020scaling, hoffmann2022chinchilla}, while effectively reducing hallucinations \cite{zhang2023hallucination} and better preserving data privacy \cite{yao2024privacy}. Early RAG systems commonly adopt a two-stage workflow: a \textit{Retrieval} stage that gathers context-relevant information from external databases, followed by a \textit{Generation} stage that incorporates the retrieved content into the prompt to produce more accurate and grounded results.

\begin{figure}[t]
    \centering
    \includegraphics[width=\linewidth]{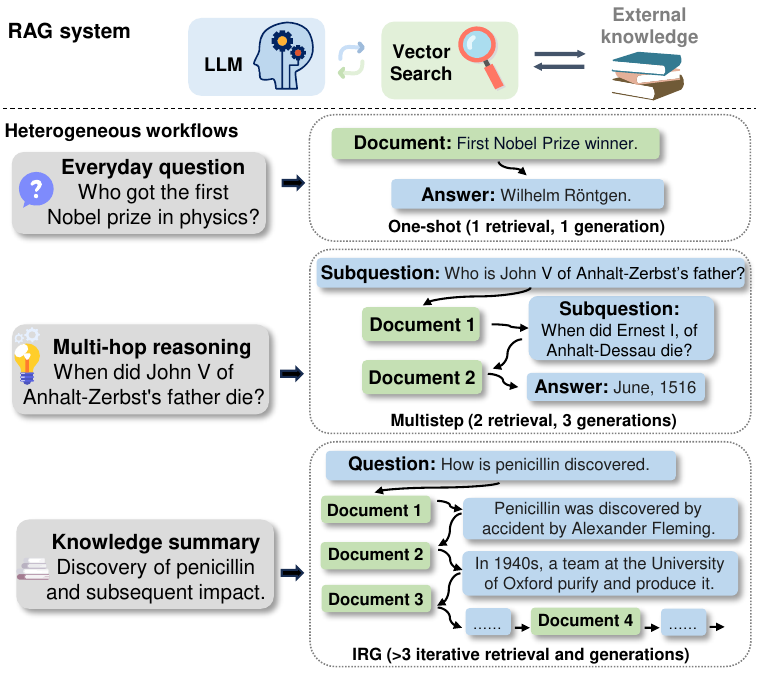}
    \caption{Heterogeneous RAG workflows bring needs for efficient LLM and vector search coordination.}
    \label{fig:intro hetero}
\end{figure}
\begin{figure*}[t]
    \centering
    \includegraphics[width=\linewidth]{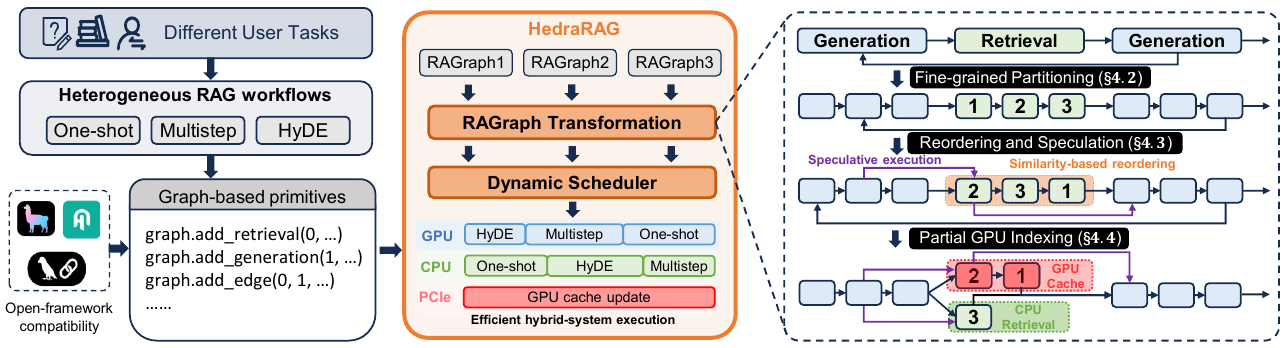}
    \caption{The overview of {\tool}, an LLM-Vector search co-designed system, to efficiently transform and schedule heterogeneous RAG workflows onto the hybrid system pipeline.}
    \label{fig:rag intro}
\end{figure*}

To support this two-stage workflow, existing systems such as LangChain \cite{langChain}, FlashRAG \cite{jin2024flashrag}, and vLLM-based pipelines \cite{kwon2023vllm} integrate LLM inference with a vector-based retrieval component through hybrid CPU–GPU designs. Typically, vector data is stored in host memory for CPU-based retrieval, while the GPU executes compute-intensive generation. 
These frameworks primarily support sequential execution of this two-stage workflow, typically combining standard LLM serving engines with vector search libraries such as FAISS \cite{douze2024faiss}, to enable basic RAG functionality.

Yet, recent advances in RAG techniques~\cite{borgeaud2022retro, shao2023IRG, asai2023selfrag, jeong2024adaptiverag, ray2024ragserve, he2025cognify, kim2023treerag, merth2024superposition, gao2022hyde, xu2024recomp} have led to increasing heterogeneity in RAG requests, both in workload patterns and workflow structures, beyond the prior simple two-stage pipeline (Figure ~\ref{fig:intro hetero}). 
This heterogeneity mainly manifests in two key dimensions. First, both the number and duration of stages can vary significantly across requests. The number of stages often increases in workflows that perform multiple iterations of retrieval and generation~\cite{borgeaud2022retro, jiang2023flare}, such as in multi-hop reasoning or refinement-based RAG. Meanwhile, the duration of each stage can fluctuate depending on factors like generation length~\cite{borgeaud2022retro}, model confidence~\cite{asai2023selfrag}, or the complexity of the input query~\cite{kim2023treerag, merth2024superposition}. Second, to support diverse objectives, different tasks often adopt distinct workflow patterns by design~\cite{jeong2024adaptiverag, ray2024ragserve, he2025cognify}. This high-level structural heterogeneity means that a general-purpose RAG framework must flexibly accommodate a wide range of task-specific workflows without requiring system reengineering.

Although existing RAG system frameworks~\cite{Liu_llamaindex_2022, langChain, jin2024flashrag} provide modular support for generation and retrieval, and work well for earlier two-stage RAG designs, they treat the two components as independently executed stages, lacking tight coordination or runtime co-optimization. As a result, dynamic and imbalanced workloads often lead to misaligned execution and CPU-GPU hybrid system under-utilization.

To mitigate these issues, we identify three key optimization opportunities for heterogeneous RAG: \textit{across stages}, \textit{within a request}, and \textit{across requests}. 
\textbf{Across stages}, generation and retrieval from independent requests can be parallelized to improve CPU–GPU pipelining. However, mismatches in execution models—step-wise, dynamic batching in LLMs versus one-shot, batched retrieval—lead to resource imbalance and frequent stalls, especially when stage lengths vary. A coordinated system design is needed to align execution and improve utilization.
\textbf{Within a request}, semantic similarity across sequential stages enables reuse and approximate retrieval. For example, embeddings in successive retrievals or partial generations often remain close to final outputs. Yet, exploiting this requires handling high-dimensional embeddings~\cite{koppen2000curse} and diverse similarity patterns, demanding solutions that balance efficiency and generality.
\textbf{Across requests}, retrievals often show skewed index access, offering GPU caching opportunities. But limited memory, high PCIe latency, and shifting access patterns make static or on-demand caching ineffective. This calls for a dynamic, runtime-aware caching strategy that adapts to evolving request workloads.

We propose \tool, a co-designed LLM–vector search system built to efficiently serve heterogeneous RAG workflows, as illustrated by Figure~\ref{fig:intro hetero}. At the core of \tool{} is \graph, a graph-based abstraction that represents diverse RAG workflows. 
\tool{} also supports seamless integration with existing open-source frameworks~\cite{Liu_llamaindex_2022, langChain, haystack} by exposing graph construction APIs compatible with their workflow specifications. 
This effectively bridges the gap between high-level workflow heterogeneity and low-level, task-oriented LLM and vector search backends.



{\graph} enables a unified view of workflow heterogeneity and facilitates runtime optimization through a set of general-purpose graph transformation operations, including node splitting, reordering, edge addition, and dependency rewiring. This abstraction unlocks a significantly larger optimization space than prior stage-centric scheduling frameworks~\cite{tan2025ayo, he2025cognify, hong2023metagpt}, by exposing finer-grained structural variations and richer dependency patterns. By carefully modeling and dynamically applying these transformations, \tool{} adapts to workload patterns and serving conditions at runtime, maximizing system throughput and resource utilization.

In particular,  \tool{} introduces three key techniques to address the system-level challenges of serving heterogeneous RAG requests:  
(1) \textit{Fine-grained sub-stage partitioning and dynamic batching} mitigate pipeline stalls caused by variable-length stages across requests, enabling smoother CPU–GPU pipelining; 
(2) \textit{Semantic-aware reordering with speculative execution} leverages intra-request similarity to overlap dependent stages in complex, multi-round workflows;  
(3) \textit{Partial GPU index caching with asynchronous updates} captures skewed access patterns across requests and enables efficient hybrid CPU–GPU retrieval execution.

\begin{figure*}[t]
    \centering
    \includegraphics[width=\linewidth]{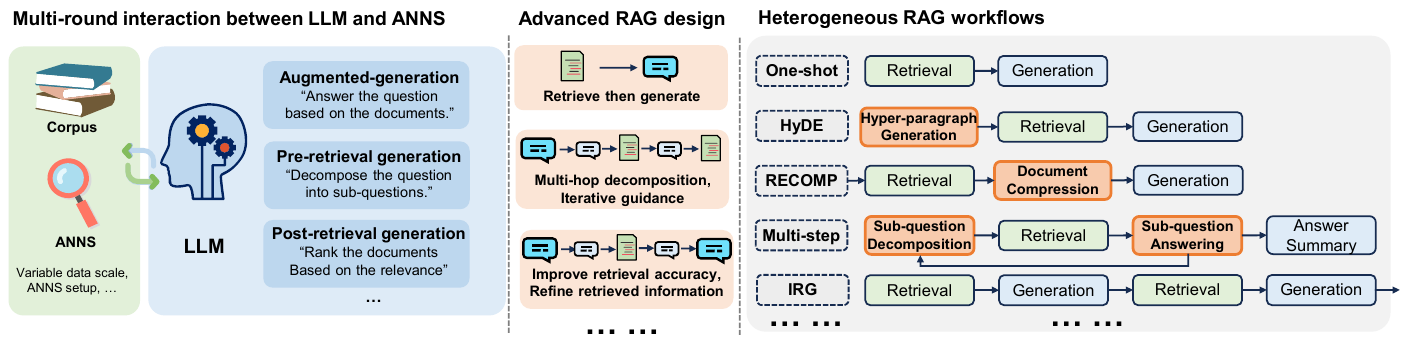}
    \caption{Advanced RAG design and heterogeneous workflows, involving multi-round LLM-ANNS interaction.}
    \label{fig:rag workflow}
\end{figure*}

The key contributions of this paper are:
\begin{itemize}
  \item We introduce a novel graph-based abstraction, {\graph}, for representing heterogeneous RAG workflows, enabling unified reasoning over diverse execution patterns on hybrid CPU–GPU platforms.
  \item We present {\tool}, a co-designed RAG serving system that leverages this abstraction to support dynamic batching, semantic-aware execution, and adaptive caching across complex multi-stage workflows.
 \item Experimental results show that {\tool} achieves over 1.5$\times$ and up to 5$\times$ throughput gains compared to state-of-the-art frameworks.
\end{itemize}

\section{Background and Related Work}

This section provides background on RAG workflows and summarizes prior work across three perspectives. Section 2.1 introduces the algorithmic roles of the retrieval and generation stages. Section 2.2 summarizes system-level work to independently optimize the retrieval and generation stages. Section 2.3 discusses recent system efforts that begin to integrate them into unified serving workflows.

\subsection{Algorithmic Roles of Retrieval and Generation}
\label{sec:workflow}

RAG workflows interleave two algorithmically distinct stages: retrieval and generation. The retrieval stage fetches relevant information—typically text fragments—from an external corpus based on a user query, while the generation stage incorporates this information into the prompt to guide LLMs toward producing responses~\cite{lewis2020rag, brown2020fewshot}.

In the simplest form, RAG follows a one-shot retrieval-then-generation pattern. However, this structure struggles to handle complex inputs that require multi-hop reasoning~\cite{ho2020wikiqa, yang2018hotpotqa} or involve vague, underspecified queries. These challenges have driven the evolution of more sophisticated algorithmic structures, resulting in heterogeneous RAG workflows, as illustrated in Figure~\ref{fig:rag workflow}.

To address retrieval difficulty, recent designs introduce \textit{pre-retrieval stages}, such as query rewriting~\cite{gao2022hyde, jiang2023longllmlingua, asai2024openscholar, peng2024taobao} and question decomposition~\cite{kim2023treerag, merth2024superposition, Liu_llamaindex_2022}, aiming to transform user inputs into more effective search queries. Conversely, to improve generation quality, \textit{post-retrieval stages} filter, rerank, or compress the retrieved content~\cite{yan2024CRAG, zhuang2024beyondyesno, yu2024rankrag, xu2024recomp} to improve contextual coherence and relevance. In parallel, the workflow has expanded beyond one-shot pattern. With increasing model reasoning capacity~\cite{openai2024reasoning, guo2025deepseekr1} and the emergence of agent-based orchestration~\cite{wang2024agentsurvey, xi2025agentsurvey2}, modern workflows incorporate multi-step generation with chain-of-thought (CoT)~\cite{wei2022cot}, verification, and feedback-guided refinement. These developments have led to branching~\cite{borgeaud2022retro, asai2023selfrag}, iterative~\cite{yue2024iterDRAG, jiang2023flare, shao2023IRG, ram2023incontextretrieval}, and adaptive~\cite{jeong2024adaptiverag, ray2024ragserve} workflow structures. 

\begin{figure}[t]
    \centering
    \begin{subfigure}{0.23\textwidth}
        \centering
        \includegraphics[width=\textwidth]{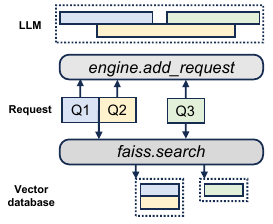}
    \end{subfigure}
    \begin{subfigure}{0.21\textwidth}
        \centering
        \includegraphics[width=\textwidth]{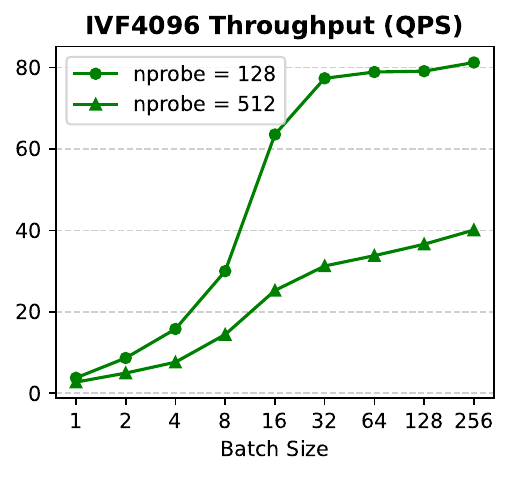}
    \end{subfigure}
    \caption{The comparison between different batching strategies: continuous batching in LLMs and fixed batching in vector search. $nprobe$ represents the number of clusters in IVF index search.}
    \label{fig:faiss_throughput}
\end{figure}

\subsection{Independent System Support}
At the system level, retrieval and generation exhibit fundamentally different hardware demands.
The \textit{generation stage} involves LLM inference, which is highly compute-intensive and runs exclusively on GPUs. Due to the auto-regressive nature of decoding~\cite{vaswani2017attention} and continuous batching across sequences of varying lengths~\cite{yu2022orca}, generation incurs dynamic workloads with significant GPU memory pressure from model weights and key–value (KV) caches.

In contrast, the \textit{retrieval stage} typically runs on CPUs, as large-scale vector indexes demand high memory capacity beyond what GPUs can support. Documents or passages are pre-encoded into semantic vectors using embedding models \cite{devlin2019bert,wang2022e5, wang2024e5multilingual}. At query time, similarity (e.g., L2 distance or cosine similarity~\cite{johnson2019similarity}) is computed between the query vector and stored vectors, returning the top-$k$ nearest matches.

To improve \textit{retrieval} efficiency, vectors are stored in Approximate Nearest Neighbor Search (ANNS) indexes. The Inverted File Index (IVF)~\cite{zobel2006ivf}, as used in FAISS~\cite{douze2024faiss}, partitions vectors into clusters via K-Means-like training~\cite{ahmed2020kmeans}, represented by centroids. At query time, the $nprobe$ closest clusters are selected, and the search is restricted to those regions, enabling a trade-off between accuracy and speed. IVF also enables spatial pruning techniques such as triangle inequality filtering~\cite{ding2015top, ding2017generalizations, xu2025tribase}.

\begin{figure*}[t]
    \centering
    \includegraphics[width=\linewidth]{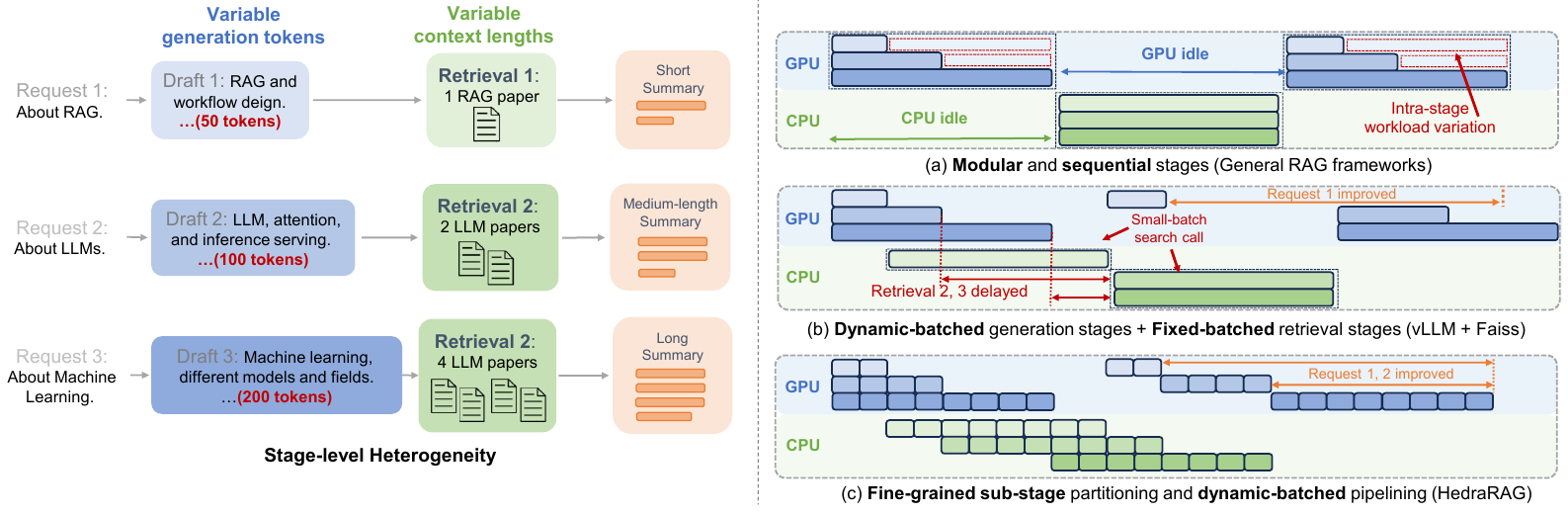}
    \caption{Comparison of CPU-GPU pipeline efficiency with different strategies.}
    \label{fig:motivation2}
\end{figure*}

\subsection{Towards Integrated Serving}
\label{sec:inte serving}
While both retrieval and generation stages are individually well-supported, recent work has begun to explore more integrated approaches for serving RAG workflows. Open-source frameworks such as LlamaIndex~\cite{Liu_llamaindex_2022}, LangChain~\cite{langChain}, and FlashRAG~\cite{jin2024flashrag} expose modular components and APIs, enabling developers to compose retrieval–generation pipelines through user-defined logic. However, these frameworks dispatch each stage to isolated backends—e.g., vLLM~\cite{kwon2023vllm} for LLM inference and Faiss~\cite{douze2024faiss} for vector search—without runtime coordination or shared optimization.

Figure~\ref{fig:faiss_throughput} illustrates the performance divergence between these two backends. vLLM maintain stable throughput via continuous token-level batching, which amortizes decoding overhead across concurrent requests. In contrast, vector search frameworks like Faiss benefit from larger batches due to multi-threaded CPU execution~\cite{dagum1998openmp}, achieving higher throughput when more queries are processed simultaneously.

Recent efforts have begun to explore optimization strategies across both system and algorithmic levels. At the system level, Chameleon~\cite{jiang2023chameleon} and RAGO~\cite{jiang2025rago} investigate resource scheduling and disaggregated deployment strategies. Yet, unified runtime support for coordinating multi-stage, heterogeneous workflows, particularly in hybrid CPU–GPU environments, remains largely absent.  At the algorithmic level, techniques such as RAGCache~\cite{jin2024ragcache}, PromptCache~\cite{gim2024promptcache}, and CacheBlend~\cite{yao2025cacheblend} aim to accelerate generation by reusing document prefixes across requests. Others including early-terminated retrieval~\cite{jin2024ragcache} and speculative generation from predicted documents~\cite{jiang2024piperag, zhang2024ralmspec}, seek to decouple retrieval latency from LLM execution. While effective in specific scenarios, these methods often rely on workflow-specific heuristics and sometimes sacrifice output quality for speed.

\section{Motivation}
\label{sec:motivation}

This section motivates our system design by identifying key performance challenges of serving heterogeneous RAG workflows in practical environments.
Although retrieval and generation are individually well-supported by existing frameworks, their composition introduces runtime bottlenecks due to stage interleaving, request variability, and resource contention.
These issues are further amplified by the dynamic and irregular structure of modern RAG workflows.
To tackle these challenges and improve overall efficiency, we identify three concrete optimization opportunities:
(1) parallelism across independent generation and retrieval stages,
(2) semantic similarity within multi-turn stages, and
(3) skewed index access across multi-request retrievals.

\subsection{Stage-Level Parallelism}
\label{sec:moti batching}
Heterogeneous RAG workflows introduce concurrent retrieval and generation stages with varying numbers and durations. A natural system-level opportunity is to pipeline these stages by integrating LLM inference and vector search through asynchronous execution. However, such naive integration suffers from mismatched execution patterns between the two components. The performance divergence between generation and retrieval backends, previously discussed in \S~\ref{sec:inte serving}, becomes more pronounced with variable-length stages.


\begin{figure}[t]
    \centering
    \includegraphics[width=\linewidth]{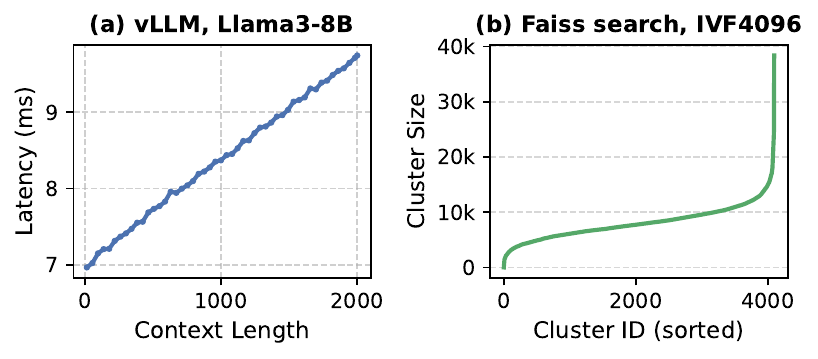}
    \caption{Workload variation in (a) LLM generation and (b) vector database query.}
    \vspace{-10pt}
    \label{fig:latency variation moti}
\end{figure}

Figure~\ref{fig:motivation2} illustrates this using a HyDE~\cite{gao2022hyde} workflow that summarizes knowledge about RAG, LLMs, and ML. In (a), general RAG frameworks execute coarse-grained stages sequentially, leading to hardware under-utilization. In (b), naive asynchronous integration introduces scheduling delays: for example, request 1’s short generation triggers an early but long-latency search, delaying retrieval for requests 2 and 3. Meanwhile, request 1's subsequent retrieval call suffers low throughput due to the small batch size. To address such inefficiency, \tool{} introduces fine-grained sub-stage partitioning for both generation and retrieval. Shown in (c), this partitioning eliminates the sequentiality of coarse-grained stages to improve throughput, and mitigates batching strategy mismatches to reduce single-request latency. 

However, achieving balanced pipelining through equal-length stage partitioning is non-trivial, due to workload imbalance across both the generation and retrieval stages. To better understand the source of imbalance, we analyze the latency characteristics of the smallest schedulable units: decoding steps for generation and single-cluster searches for retrieval. As shown in Figure~\ref{fig:latency variation moti}, the latency distributions of both decoding steps and retrieval clusters are highly non-uniform, further complicating static partitioning. To mitigate such imbalance, we design a dynamic, load-aware alignment strategy that adjusts sub-stage boundaries based on real-time workload, enabling efficient and stall-free hybrid pipelining.



\subsection{Intra-Request Semantic Similarity}
\label{sec:moti-similarity}

\begin{figure}[t]
    \centering
    \includegraphics[width=\linewidth]{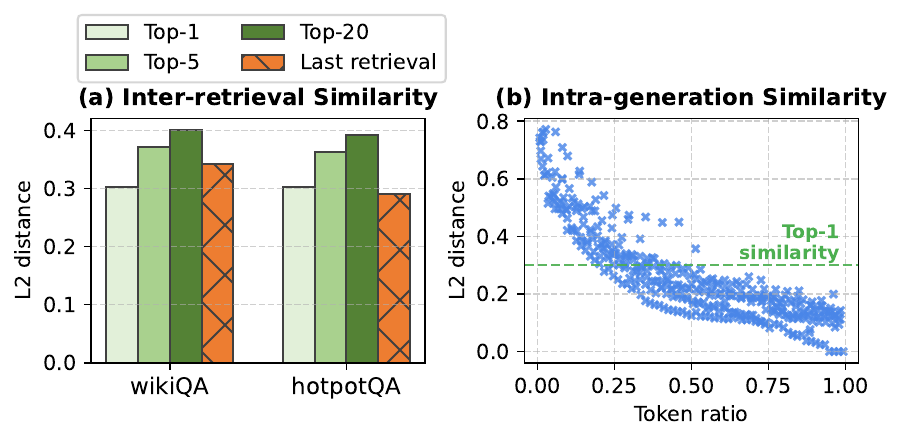}
    \caption{Inter-stage and Intra-stage similarity in IRG. (a) The distances of the current query retrieval vector to its top-$k$ retrieved passage embedding vectors, and to the previous retrieval query vector. (b) The embedding vector distances between partial generations with different prefix ratios and the final generation result.}
    \label{fig:motivation1}
\end{figure}

Heterogeneous RAG workflows often involve multi-round generation and retrieval~\cite{Liu_llamaindex_2022, shao2023IRG}.
However, existing systems typically treat these stages as independent requests. Such separation results in accumulated latency and redundant computation, while missing opportunities for cross-stage coordination and optimization.

Our key observation is that semantic similarities naturally emerge across stages within the same request. These arise from the inherent coherence of generation and retrieval in language workflows. We identify two types of such similarity, illustrated through an IRG~\cite{shao2023IRG} workflow on open datasets:

\noindent \textbf{(1) Inter-retrieval similarity:} The similarity between query embedding vectors in adjacent retrieval stages is often high, since these queries are usually generated from the same underlying context.
As shown in Figure~\ref{fig:motivation1}(a), the average distance between consecutive queries is smaller than between a query and its top-5 retrieved passages. This suggests that successive retrievals operate within similar index regions and may even yield overlapping results. 

\noindent \textbf{(2) Intra-generation similarity:} the step-wise nature of LLM decoding leads to partial generations that are semantically close to the final output. As Figure~\ref{fig:motivation1}(b) shows, using only 22–50\% of the tokens yields embeddings within the top-1 retrieved range of the final output, indicating strong semantic proximity.

These observations unlock opportunities for intra-request optimization. Instead of adhering to strict sequential execution across stages in iterative workflows, we can leverage such semantic locality to reduce the costs of both generation and retrieval.
However, traditional locality-based pruning methods~\cite{ding2015top, ding2017generalizations} are less effective in high-dimensional embedding spaces. To this end, our second design leverages heuristic, semantic-aware optimizations: we introduce locality-based cluster reordering and workload-aware speculative execution to overlap dependent generation and retrieval stages, improving latency without sacrificing result quality.

\subsection{Inter-Request Retrieval Skewness}
\label{sec:skewness}

When dealing with large external databases, the highly concurrent retrieval stages in heterogeneous RAG workflows often leads to system-level bottlenecks, incurring higher overhead than generation. One potential solution is to offload expensive vector similarity computations to the GPU. However, existing GPU-accelerated vector search engines~\cite{johnson2019faissgpu, johnson2019searchgpu, wang2021milvus, zhang2024fastreorder} are designed for standalone use and do not address co-execution challenges with LLM inference.


In RAG serving on hybrid CPU–GPU environments, limited GPU memory poses a fundamental challenge. LLM inference consumes most of the GPU memory for model weights and KV caches, leaving insufficient space to store the full vector index. As a result, systems must load index shards on demand from CPU memory. However, this is bottlenecked by the limited bandwidth of PCIe~\cite{neugebauer2018understandingPCIE}, making such transfers prohibitively expensive at runtime.

\begin{figure}[t]
    \centering
    \includegraphics[width=\linewidth]{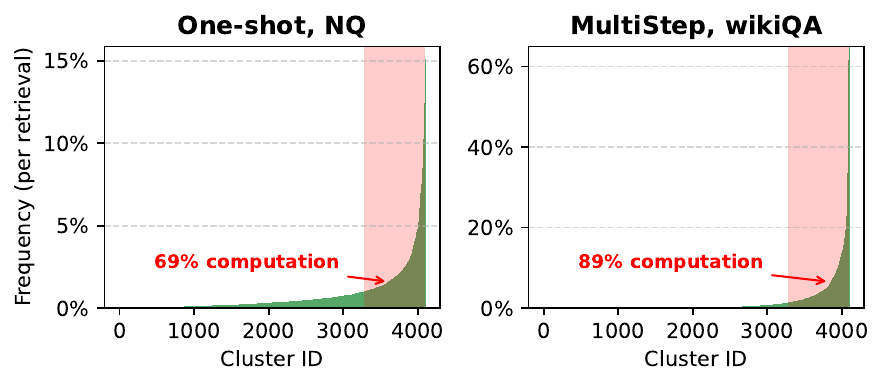}

    \caption{Skewness of access frequency among clusters when using IVF4096 indexing. Cluster IDs are sorted in descending order of frequency.}
    \vspace{-10pt}
    \label{fig:skewness}
\end{figure}

To address this, we leverage an important workload characteristic: index access skewness. Real workloads exhibit concentrated access to a small subset of index clusters, which we refer to as \textit{hotspot clusters}. This skewness arises because user requests centered around similar topics or scopes tend to generate query embeddings spatially close to a shared subset of clusters in the index. As shown in Figure~\ref{fig:skewness}, the top 20\% of hotspot clusters account for over 69\% of total computation. This suggests that caching only a fraction of the index could yield substantial acceleration.

However, hotspot clusters shift dynamically across heterogeneous workflows and request distributions. We therefore introduce a partial GPU index cache with asynchronous updates. This design enables lightweight, runtime-aware caching of hot clusters, allowing high-throughput GPU search while minimizing interference with ongoing LLM generation and CPU-side retrieval.

\section{{\tool}: Method and Design}
We present {\tool}, a co-designed framework for LLM and vector search integration, built to efficiently serve heterogeneous RAG requests on a CPU-GPU hybrid system. {\tool} abstracts user-defined RAG workflows as a graph-based abstraction, and enables unified optimization techniques through a series of graph transformation operations. 
{\tool} extensively explores optimization opportunities across the stage parallelism, intra-request semantic similarity, and inter-request retrieval skewness, and encapsulates the optimization techniques as graph transformations, including node splitting, reordering, edge addition, and dependency rewiring. Through dynamic graph transformation and scheduling, {\tool} efficiently coordinates and parallelizes stages across concurrent, heterogeneous requests. Such design bridges the gap between highly variable runtime workflows and the underlying LLM and vector search backends, enabling robust and generalizable optimization.

\definecolor{lightblue}{RGB}{173, 216, 230}  
\definecolor{lightgreen}{RGB}{144, 238, 144} 

\subsection{{\graph}: RAG Specific Abstraction}

We first introduce {\graph}, a graph-based abstraction tailored for RAG workflows, to enable customizable workflow specification and optimization. 
The original {\graph} consists of two types of nodes: {\textbf{Generation}} nodes initiate LLM generation with the specific prompt and inputs,
while {\textbf{Retrieval}} nodes perform vector database search to fetch relevant passages. {\tool} provides simple graph primitives that allow users to construct {\graph} with the target RAG workflows, as is shown in Listing \ref{list:dsl}. Through node adding primitives \texttt{add\_generation} and \texttt{add\_retrieval}, the user can define Generation or Retrieval nodes with a customized prompt, typically corresponding to a single stage in the RAG workflow.
\texttt{add\_edge} connects different stages by establishing edges, enabling data flow and control transitions, including conditional branches. Such a design is compatible with many existing RAG frameworks \cite{langChain, haystack}, thereby enabling seamless migration and integration.

Unlike vanilla directed-graph-based workflow representations \cite{onnx, paszke2019pytorch, tan2025ayo}, {\graph} captures the execution asymmetry between retrieval and generation stages. Each Retrieval node performs a predefined sequence of searches over a fixed subset of the index clusters, with structurally bounded execution. In contrast, each Generation node represents prompt-based LLM inference, realized as a dynamic, multi-step process that gradually unfolds at token level.

{\tool} further elevates system-level optimization by formulating it as a set of graph transformation operators over the {\graph} abstraction. These transformations, including node splitting, reordering, and edge insertion, convert the original stage-wise, sequential workflow into fine-grained, overlappable sub-stages, enabling dynamic and parallel scheduling across hybrid hardware resources. Such a graph-based abstraction overcomes the limitations of existing ad-hoc workflow optimization approaches \cite{zhang2024ralmspec, jiang2024piperag, jin2024ragcache}, providing a unified and generalizable system-level optimization framework for heterogeneous RAG workflows.

\definecolor{lightgray}{gray}{0.97}
\definecolor{codegray}{gray}{0.4}
\definecolor{codeblue}{rgb}{0.25,0.35,0.75}


\definecolor{codegreen}{rgb}{0,0.6,0}
\definecolor{codegray}{rgb}{0.5,0.5,0.5}
\definecolor{codepurple}{rgb}{0.58,0,0.82}
\definecolor{backcolour}{rgb}{0.95,0.95,0.92}
\definecolor{textblue}{rgb}{.2,.2,.7}
\definecolor{textred}{rgb}{0.54,0,0}
\definecolor{textgreen}{rgb}{0,0.43,0}
\definecolor{codered}{rgb}{201,72,12}

\lstdefinestyle{ragdsl}{
language=Python,
basicstyle=\linespread{1}\ttfamily\footnotesize,
breaklines=true,
numbers=left,
frame=single,
numberstyle=\tiny, 
stepnumber=1,
numbersep=5pt, 
tabsize=4,
keywordstyle=\bfseries\color{codegreen},
commentstyle=\color{textred},   
stringstyle=\color{textgreen},
columns=fullflexible,
keepspaces=true,
xleftmargin=\parindent,
showstringspaces=false,
otherkeywords = {True, False},
keywordstyle=[2]\color{codepurple}\bfseries,
keywords=[2]{GNNAdvisor, GNNA},
keywordstyle=[3]\color{textblue}\bfseries,
keywords=[3]{__init__, forward},
keywordstyle=[4]\color{codegreen},
keywords=[4]{self},
}

\begin{figure}[t]
\begin{minipage}{\linewidth}
\begin{lstlisting}[style=ragdsl, caption={Construct RAG workflows with graph primitives.}, label={list:dsl}]
from HedraRAG import RAGraph, START, END
from HedraRAG import Server
# HyDE-style workflow
g1 = RAGraph()
g1.add_generation(0, prompt="Generate a hypothesis 
                for {input}.", output="hypopara")
g1.add_retrieval(1, topk=5, query="hypopara", output="docs")
g1.add_generation(2, prompt="Answer {query} using {docs}.")
g1.add_edge(START, 0); g1.add_edge(0, 1)
g1.add_edge(1, 2); g1.add_edge(2, END)
# Multistep-style workflow
g2 = RAGraph()
g2.add_generation(0, prompt="Decompose {input} into 
                subquestions.", output="subquestion")
g2.add_retrieval(1, topk=2, query="subquestion", 
                 output="docs")
g2.add_generation(2, prompt="Answer {subquestion} 
                using {docs}.")
g2.add_edge(START, 0); g2.add_edge(0, 1); g2.add_edge(1, 2)
g2.add_edge(2, lambda s: 1 if s.get("subquestion") else END)
# Server initiating and execution
s = Server(generator="Llama3-8B", index="IVF4096")
s.add_request("What is RAG?", g1)
s.add_request("Compare RAG with long-context models.", g2)
\end{lstlisting}
\end{minipage}
\end{figure}

\subsection{Fine-Grained Sub-Stage Pipelining}
\label{sec:partition}

To bridge the design gaps between LLM generation and vector search, {\tool} partitions both generation and retrieval stages into fine-grained sub-stages with similar execution costs. In generation, each sub-stage comprises several decoding steps. In retrieval, each sub-stage involves searching across one or more clusters. Such partitioning follows two key objectives: (1) By aligning the short-latency, multi-step decoding with the long-latency, single-step retrievals, we can enable coordinated dynamic batching across generation and retrieval stages. 
(2) these sub-stages serve as the fundamental units for scheduling and execution, with each representing a portion of a generation/retrieval stage's workload. Such design enables further optimizations including speculative execution and partial GPU indexing.


A straightforward method to partition sub-stages is to assign a fixed number of generation steps and retrieval clusters. However, such method leads to sub-stage misalignment and workload imbalance, as both LLM generation steps and single-cluster retrieval operations exhibit runtime workload variation. To overcome this, we introduce a dynamic time-budgeting method based on retrieval requests. Before executing a sub-stage, clusters from each retrieval request are incrementally added until a maximum time budget $mb$ is reached. The execution time for the sub-stage is then determined as the time cost to  batch-search these clusters. The retrieval-centric strategy is motivated by the fact that the workload variance across retrieval clusters is substantially higher than that of generation steps.

The configuration of $mb$ is crucial to performance, involving the tradeoff between the latency improvement of sub-stages and the additional overhead introduced by partitioning and scheduling. We calculate $mb$ by modeling the expected latency improvement $\Delta_l$:
\begin{equation}
    mb=\text{argmax}(\Delta_l), \\ {\Delta_l} = \frac{t_\text{\textbf{Retrieval}} - mb}{2} + \frac{t_\text{\textbf{Retrieval}}}{mb}\beta,
\end{equation}
where $\beta$ denotes the CPU overhead of request scheduling and handling intermediate results. $t_\text{\textbf{Retrieval}}$ denotes the average time of retrieval stages, measured at runtime. In the equation, we assume that retrieval requests arrive evenly across all sub-stages, so the expected wait time for the preceding retrieval operation is reduced from $\frac{t_\text{\textbf{Retrieval}}}{2}$ to $\frac{mb}{2}$.

In {\graph}, sub-stage partitioning is modeled via node splitting, where coarse-grained nodes are divided into fine-grained, sequentially dependent sub-nodes with similar costs. Efficient CPU-GPU pipelining is enabled by concurrently scheduling these fine-grained nodes across different requests.

\subsection{Similarity-Aware Search Optimization}
\label{sec:speculation}

To further reduce the latency of RAG requests involving multi-round generation and retrieval stages, {\tool} leverages the intra-request the semantic similarity. We first define the optimization problem for similarity-based vector search as follows: given two query vectors $v$ and $v'$, with cluster sets $C=\{c_1, c_2, ..., c_{nprobe}\}$ and $C'=\{c_1', c_2', ..., c_{nprobe}'\}$ to search. Assuming their distance satisfies $d_{vv'} \le \delta$, how to leverage the search results of $v$ to accelerate the search for $v'$?

Leveraging such semantic vector similarity is challenging due to the well-known curse of dimensionality \cite{koppen2000curse}. Existing semantic embedding models produce vectors in high-dimensional space (e.g. 768 for BERT \cite{devlin2019bert} and 1024 for e5\_large \cite{wang2022e5}), leading to sparse distributions on the surface of spheres. As a result, pairwise distances tend to become nearly uniform. Traditional similarity-based optimizations, such as those using triangle inequalities \cite{ding2015top, ding2017generalizations, xu2025tribase}, become significantly less effective.

\begin{figure}[t]
    \centering
    \includegraphics[width=\linewidth]{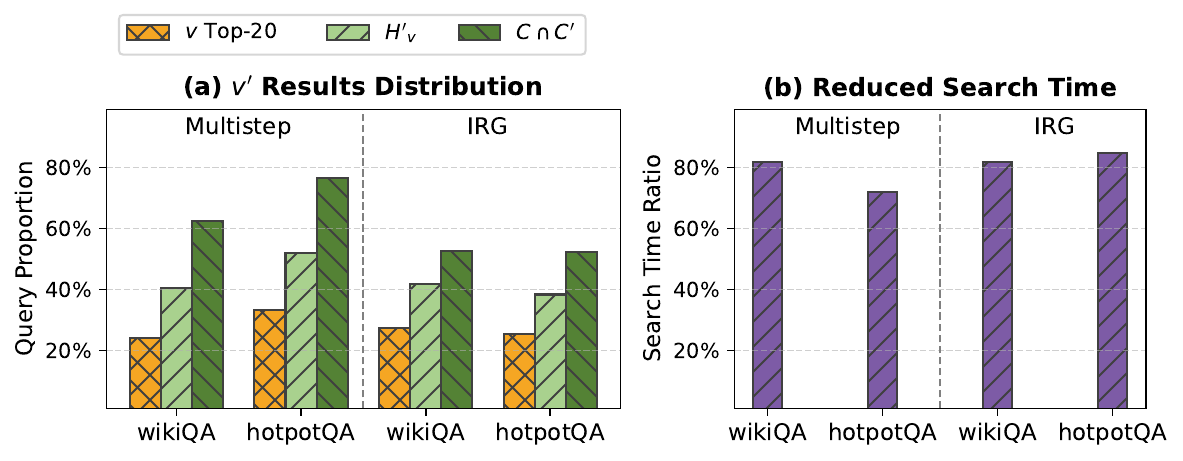}
    \caption{Opportunities from semantic similarity in RAG. (a) Proportion of queries satisfying each locality-based observation, where $v'$ top-$k$ is set as 2. (b) The reduction of effective search time throught locality-based reordering. }
    \label{fig:similarity opportunity}
\end{figure}

Fortunately, we find that semantic similarity still provides good opportunities specific in the RAG context. Our experimental analysis reveals three \emph{locality-based observations} related to semantic similarity: (1) The search results of $v'$ tend to be included within the search results of $v$ with a larger top-$k$. (2) When the search results of $v$ are in a cluster set $H_v$, the results of $v'$ also tend to be located in $H_v$. (3)  The search results of $v'$ tend to be located in clusters of $C\cap C'$. According to the open-domain dataset results shown in Figure~\ref{fig:similarity opportunity}(a), up to 33\%, 52\%, and 77\% of retrieval queries satisfy the three locality-based observations, respectively.

{\tool} leverages the above observations by caching and reusing historical search information. For each retrieval in a request, a set of larger top-$k$ results of $v$ (20 in practice) are stored in a local cache for future reuse. The search for $v'$ is first attempted in the local cache of $v$. Since the primary cost of retrieval lies in computing vector similarity, maintaining and querying the local cache incurs negligible additional overhead. 
Next, the target cluster set $C'$ is reordered based on $H_v$ and $C_v$: first search $H_v \cap C'$ (if any), followed by $(C - H_v) \cap C'$ (if any), and finally the remaining clusters. As illustrated in Figure \ref{fig:similarity opportunity}(b), such search order optimization leads to earlier termination in ANNS by up to 28\%, thereby reducing the effective search time.

\begin{figure}[t]
    \centering
    \includegraphics[width=\linewidth]{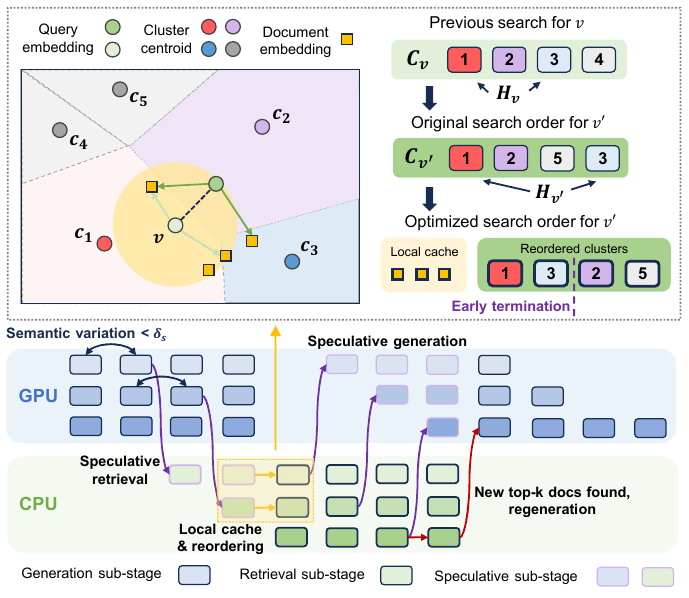}
    \caption{Leveraging semantic similarity with speculative generation and retrieval sub-stages.}
    \label{fig:speculation}
\end{figure}

Based on the search order optimization, {\tool} further exploits the early-terminating property of ANNS to enable speculative execution. As illustrated in Figure \ref{fig:speculation}. we enable two forms of speculative execution: When a retrieval stage is followed by a generation stage, \textit{speculative generation} can be started using partial search results from a small subset of clusters. This allows the following generation stage, which would otherwise run sequentially only after the entire retrieval stage completes, to overlap with the remaining cluster searches. After the retrieval stage completes, the final search results are compared against the partial results used by speculative generation. If the results are identical, the speculative generation is valid, 
otherwise the generation stage must be restarted. Since speculative generation overlaps with ongoing retrieval, re-generation does not add to the originally expected latency.

When a generation stage is followed by a retrieval stage, \textit{speculative retrieval} can be initiated using embeddings from partially generated outputs. Speculative retrieval is pipelined in parallel with the remaining generation stage, to provide inter-retrieval history to guide the following real retrieval stage. This optimization is effective when no preceding retrieval is available, or the generation between adjacent retrieval stages is long.

Deciding the speculative execution point in heterogeneous RAG workflows is challenging due to unpredictable outputs and workflow patterns. Most existing approaches \cite{zhang2024ralmspec, jiang2024piperag, jin2024ragcache} are tailored to specific workflows and depend on static heuristics, with limitations to generalize and adapt under dynamic serving senarios. \tool{} introduces an adaptive speculative strategy based on both workload dynamics and semantic similarity. Specifically, speculative generation/retrieval is triggered when CPU/GPU system throughput of the next sub-stage is underutilized with $\frac{T_{\text{curr}}}{T_{\text{max}}} < \tau$, where $T_{\text{curr}}$ is empirically estimated with the number of requests and prefill tokens. $T_{\text{max}}$ is the estimated system peak throughput. 

For each sub-stage, {\tool} selects speculative requests until system throughput reaches threshold $\tau$ at each sub-stage. For speculative generation, we select the retrieval request with current top-$k$ vectors closest to the query embedding. For speculative retrieval, we select the generation stage with minimal semantic drift $\delta_s$ from the previous sub-stage. Such strategies heuristically prioritizes those with lower speculative error rates.

In {\graph}, search order optimization and speculative execution modify the dependency structure between sub-nodes. Leveraging semantic similarity, the sub-nodes obtained are reordered for an optimized execution sequence. Speculative edges are inserted to mark the entry points of speculative execution, enabling overlapping between originally sequential sub-stages and supporting rollback upon speculation errors.

\subsection{Partial GPU Indexing}
\label{sec:onloading}

Leveraging the cluster skewness described in \S\ref{sec:skewness}, we can only cache a small number of hotspot clusters to accelerate most of the computations. 
{\tool} further utilizes this observation to build a hybrid retrieval engine with a partial index cache for index clusters, enabling cross-device acceleration of vector search.

\begin{figure}[t]
    \centering
    \includegraphics[width=\linewidth]{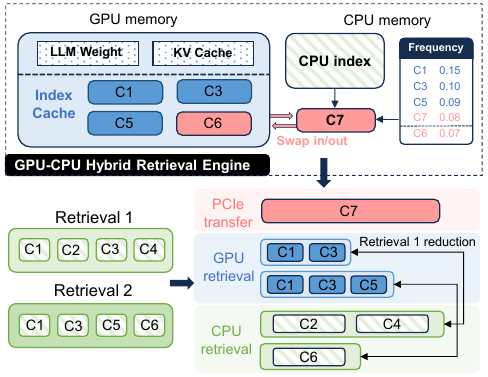}
    \caption{Hybrid retrieval engine of {\tool} with $gc=4$. $C_i$ represents the $i$-th cluster.}
    \label{fig:hybrid_retrieval}
\end{figure}

To efficiently identify and onload hotspot clusters, {\tool} maintains a GPU-side partial index cache for the CPU index. As illustrated in Figure \ref{fig:hybrid_retrieval}, {\tool} allocates a certain amount of GPU memory for the cache, and tracks the runtime access frequency of each cluster in the index. The top-$gc$ most frequently accessed clusters are retained in the cache. Cache updates primarily involve swapping clusters in and out via asynchronous memory transfers, which are executed in parallel with the ongoing stages on both the CPU and GPU. To avoid PCIe contention caused by overly frequent cache updates, {\tool} performs updates at fixed intervals, set as every 50 sub-stages in practice.

For each batched retrieval workload within a sub-stage, the hybrid retrieval engine first checks whether each target cluster resides in the GPU cache.
If the cluster is valid in the partial index cache, the search is performed on the GPU. If the cluster is not in the cache or is currently being swapped in or out, the search is performed on the CPU. All the clusters assigned to GPU and CPU computation are batched through unified search interfaces (\S \ref{sec:implementation}), enabling efficient thread-level parallelism. After the search calls completes, the results from both CPU and GPU for each request are merged. By such design, {\tool} enables the parallelism of PCIe transfers, CPU-side retrieval, and GPU-side retrieval, thereby maximizing hardware utilization to improve retrieval efficiency.

The co-location of LLM weights and KV cache necessitates a trade-off in setting the GPU index cache size ($gs$): too small yields minimal retrieval benefit, too large interferes with generation via KV cache swapping. To address this, {\tool} conducts offline benchmarking on open datasets to characterize the generation throughput $T_G(KV\_size, rps)$ and the CPU retrieval throughput $T_R(rps)$ under varying request rates. For each new RAG workflow, {\tool} estimates the expected generation and retrieval request rate ($rps_G$, $rps_R$) from its average stage composition, and selects the KV cache size by solving:
\begin{equation}
\arg\max_{KV\_size} \min\left\{T_G(\mathsf{KV\_size}, rps_G),\ T_R(rps_R)\right\}.
\end{equation}
When the server starts, we allocate $KV\_size$ GPU memory to the KV cache, and the remaining GPU memory budget is allocated for caching index clusters.

In {\graph}, GPU indexing is modeled as further parallelization within a sub-stage over its assigned clusters. For each sub-stage, {\tool} decides whether to enable GPU acceleration based on the number of target clusters cached on the GPU, balancing the potential search speedup against kernel launch and synchronization overhead.

\subsection{Dynamic Graph Transformation and Scheduling}
\label{sec:dynamic}

The heterogeneity of RAG workflows results in distinct workload distributions for the generation and retrieval stages. Consequently, the effectiveness of each optimization technique may vary with the workflow type and the runtime workload. Therefore, {\tool} introduces adaptive graph transformation and scheduling.

\begin{figure*}[t]
    \centering
    \includegraphics[width=\linewidth]{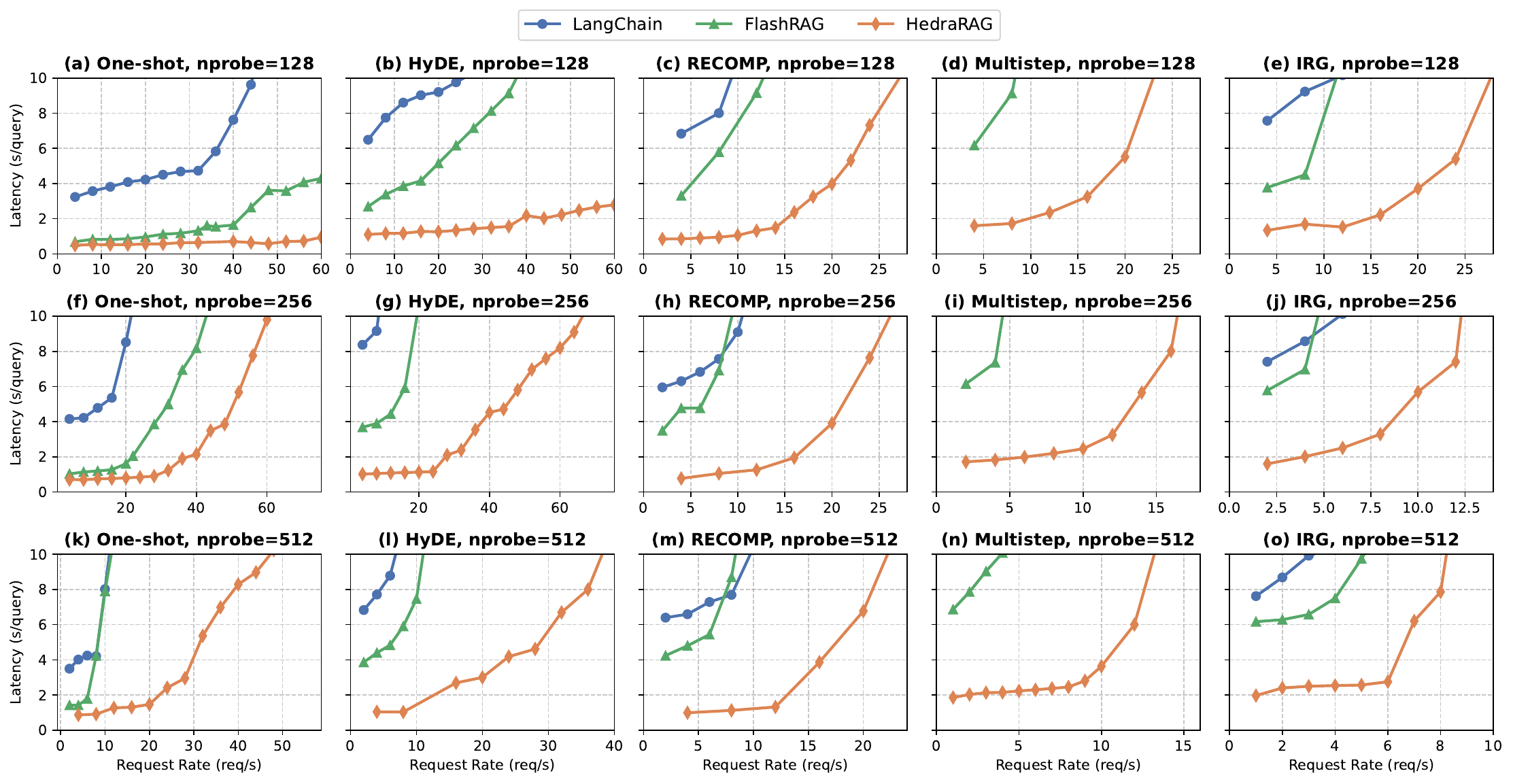}
    \caption{Average request latency when using various RAG workflows.}
    \label{fig:first_latency}
\end{figure*}

Before each scheduling cycle, the scheduler traverses a batch of pending requests to identify stages that can be executed in parallel, forming a node wavefront. 
The scheduler then performs graph transformations sequentially, guided by the estimated latency and throughput benefits. The resulting optimized sub-stages are then dispatched to the CPU-GPU execution pipeline, providing foundational support for coordinated optimization of heterogeneous RAG workflows.

In addition, {\graph} enables the incorporate a broader range of optimization opportunities. By defining graph transformation operations along with their expected latency shifts, various existing workflow optimizations can be naturally integrated. For example, retrieval-generation workflow optimization \cite{jiang2024piperag, jin2024ragcache, zhang2024ralmspec} can be modeled by introducing new speculative edge, and the GPU index prefetching \cite{lin2025telerag} can be implemented by introducing specific nodes to perform onloading that execute in parallel with retrieval nodes.

\section{Implementation}
\label{sec:implementation}
\noindent \textbf{System Construction}. {\tool} is built on vLLM \cite{kwon2023vllm} (version 0.6.6) and Faiss \cite{douze2024faiss} (version 1.9.0). The generation and retrieval workers are assigned to separate processes using Python multiprocessing, enabling parallel execution. At runtime, the generation worker repeatedly invokes the \emph{step} function of the vLLM engine, while the retrieval worker executes the extended \emph{step} function of index search. The two workers exchange inputs and outputs via a shared message queue.
During each parallel iteration cycle, the scheduler traverses the {\graph} of all active requests, selects a new wavefront, and inserts the transformed sub-nodes into the corresponding task queues of the generation and retrieval workers.

\noindent \textbf{Extension of Vector Search Library}. Faiss provides a state-of-the-art in-memory vector search implementation with various performance optimizations. However, its interface is primarily designed for batched, search-only operations, making it difficult to integrate with the fine-grained execution patern of LLMs.
We extend the index searching implementation with multi-step cluster partitioning and step-wise execution, and provide the \emph{step} function similar to LLM engines.
We also implement an interface for asynchronous index loading, supporting partial GPU caching of selected clusters.
Furthermore, since \#clusters to be processed per batch can vary within a sub-stage at both the CPU and GPU side, we provide a variable-length cluster search interface across requests, along with specific performance optimizations including workload balancing and effective reduction.

\section{Evaluation}

\subsection{Experimental Setup}

\textbf{Hardware}.
We evaluate {\tool} on a CPU–GPU hybrid server. Vector search is executed on an AMD EPYC 9534 64-core processor, while LLM generation is performed on NVIDIA H100 GPUs with 80 GB of memory.

\noindent \textbf{Model}. We use Llama 3.1–8B~\cite{dubey2024llama} as the primary model in our experiments. Additionally, we also evaluate {\tool} on Llama2-13B \cite{touvron2023llama2} and OPT-30B \cite{zhang2022opt}. We use the instruct-tuned model version to better construct RAG workflows.

%
%
\noindent \textbf{Corpus and Index}. Wikipedia passages~\cite{chen2017wiki, izacard2023atlas} as the primary retrieval corpus, covering knowledge up to 2022 and containing $\sim$38M documents. 
%
For each document chunk in the corpus, we use the e5\_large embedding model \cite{wang2024e5multilingual} for 1024-dimensional semantic vectors. We evaluate the index commonly used for large-scale vector search: IVF4096. $nprobe$ is set to 128, 256, or 512, for different search costs and accuracy. The number of top-$k$ results returned is 1.

\vspace{1pt}
\noindent \textbf{Datasets}. We evaluate on three open datasets: NaturalQuestions \cite{kwiatkowski2019naturalqa} (referred to as NQ), 2WikiMultiHopQA \cite{ho2020wikiqa} (referred to as wikiQA), and HotpotQA \cite{yang2018hotpotqa}. Both wikiQA and HotpotQA are designed for multi-hop or progressive question answering, which is typically challenging for the simple one-shot RAG workflow.

\begin{figure}[t]
    \centering
    \includegraphics[width=\linewidth]{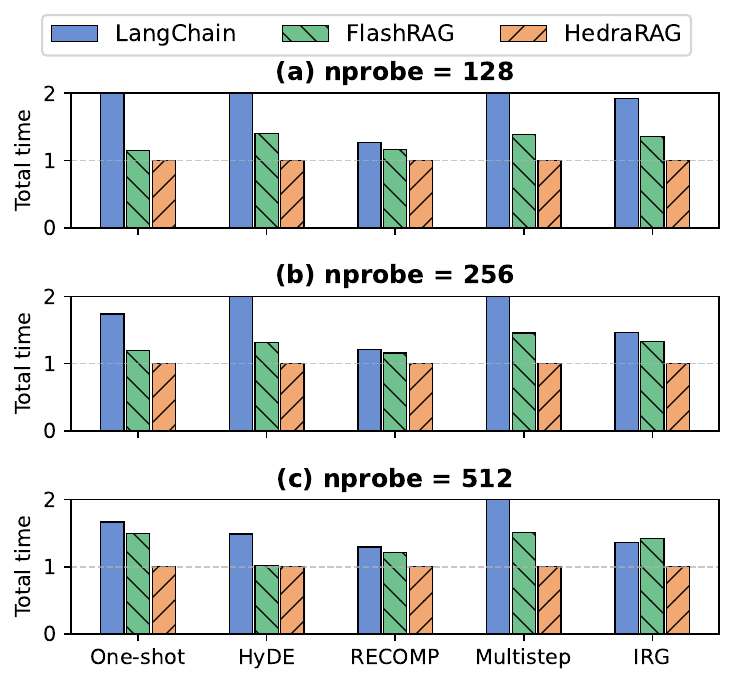}
    \caption{Total offline runtime of different RAG frameworks. All timing results are reported with normalization w.r.t. {\tool}. For clarity, values over 2 are truncated.}
    \label{fig:offline_latency}
\end{figure}

%
We evaluate five types of RAG workflows: One-shot, Multistep \cite{Liu_llamaindex_2022}, IRG \cite{shao2023IRG}, HyDE \cite{gao2022hyde}, and RECOMP \cite{xu2024recomp}. One-shot is the simplest retrieval-then-generation workflow. Multistep and IRG involve multiple rounds of interaction between generation and retrieval stages, while HyDE and RECOMP introduce additional pre-retrieval and post-retrieval stages, respectively.

\noindent \textbf{Baselines}. Our baseline includes two open-source RAG frameworks: LangChain \cite{langChain} and FlashRAG \cite{jin2024flashrag}. The frameworks support comprehensive functionality for various RAG workflows, and provide implementations that integrate with the state-of-the-art inference serving system (vLLM~\cite{kwon2023vllm}) and the vector database search library (Faiss~\cite{douze2024faiss}).
%
%
For speculative execution, we compare two existing approaches: RaLMSpec \cite{zhang2024ralmspec} and RAGCache \cite{jin2024ragcache}. As neither of them provides open-source access, we enable support for both in {\tool} by adding speculative execution edges between the generation and retrieval nodes.

\noindent \textbf{Primary Evaluation Setting.} To facilitate a detailed and systematic comparison of different RAG workflows, we primarily conduct in-depth evaluations on LLaMA3-8B~\cite{dubey2024llama}, focusing on the impact of retrieval-stage overheads (e.g., varying $nprobe$ from 128 to 512) and workflow patterns (5 RAG workflows). This setup enables us to thoroughly analyze performance bottlenecks and optimization effectiveness in a controlled setting. To further validate the generality of our findings, we additionally evaluate {\tool} on larger models (e.g., LLaMA2-13B~\cite{touvron2023llama2}, OPT-30B~\cite{zhang2022opt}), and observe consistent performance improvement across these settings.

\subsection{Overall Improvement}

We evaluate {\tool} under three scenarios: single-workflow online serving, offline execution, and multi-workflow concurrent serving, measuring its impact on system latency and throughput. We set the service-level objective (SLO) to 10 seconds per request, representing the target for end-to-end responsiveness to user queries.

\noindent \textbf{Online serving}. We first evaluate how {\tool} improves throughput and latency in online RAG request serving. Figure \ref{fig:first_latency} illustrates how request latency varies with the request arrival rate across different RAG workflows and datasets. Compared to existing RAG frameworks, {\tool} reduces request latency by 2.2$\times$ to as much as 18.2$\times$ at the same request rate. {\tool} also sustains higher request rates, achieving more than 3$\times$.
Performance gains stem from {\tool}’s efficient parallelization of the generation and retrieval stages, along with associated optimization strategies.


Through performance variation of vertical subplots in Figure \ref{fig:first_latency}, we can observe that {\tool} provides greater improvements on more complex workflows. For instance, when $nprobe = 256$, the throughput improvement on one-shot is 1.5$\times$, and reaches up to 4$\times$ and 3$\times$ for Multistep and IRG. Moreover, the performance variation across the horizontal subplots in Figure~\ref{fig:first_latency} reveals how the retrieval-stage overhead influences the optimization effectiveness of {\tool}. For example, in the one-shot workflow, increasing $nprobe$ from 256 to 512 leads to a throughput improvement from 1.5$\times$ to 4.4$\times$. This is because {\tool}’s fine-grained, dynamic graph transformation and scheduling mechanism significantly reduces pipeline stalls caused by vector search, and further improves the efficiency of multi-round interactions between the LLM and vector search. 


\begin{figure}[t]
    \centering
    \includegraphics[width=\linewidth]{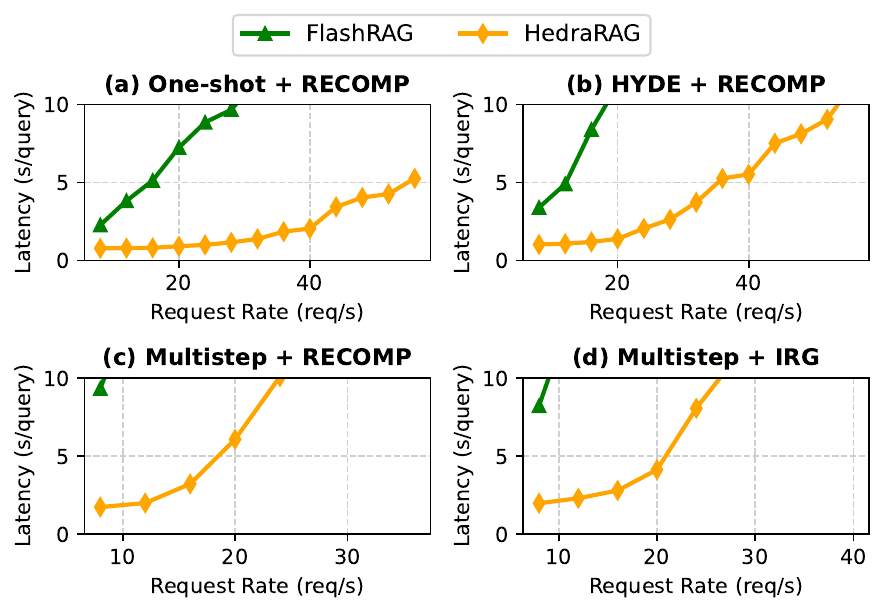}
    \caption{Average request latency with concurrency of different RAG workflows. $\mathit{nprobe}$ is set as 128.}
    \label{fig:hete_latency}
\end{figure}

\begin{figure}[t]
    \centering
    \includegraphics[width=\linewidth]{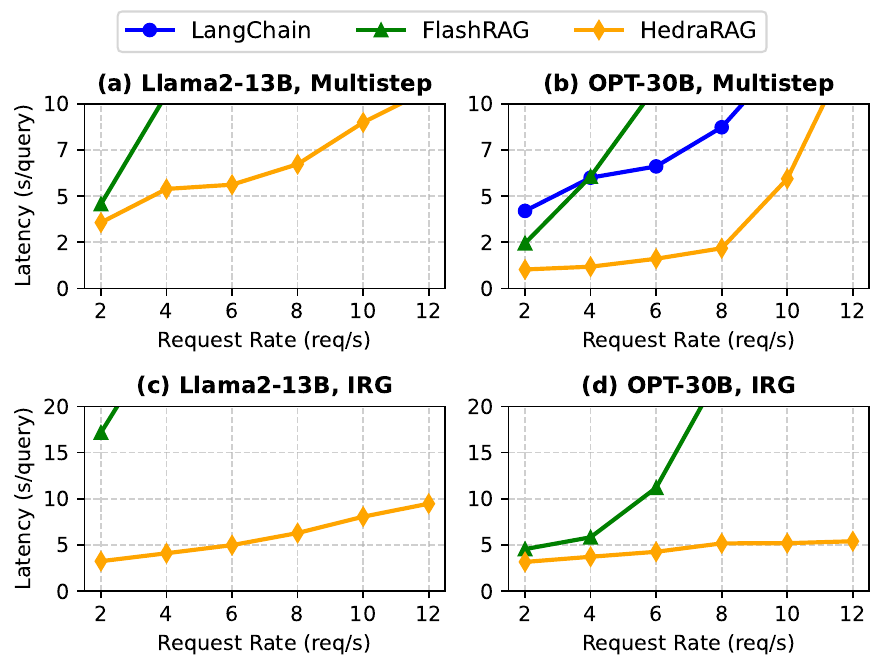}
    \caption{Average request latency on larger LLM models, with Multistep and IRG workflows. $nprobe$ is set as 128.}
    \label{fig:other_model}
\end{figure}

\vspace{0.8em}

\noindent \textbf{Offline execution}. Next, we evaluate how {\tool} improves the execution time for offline workload. Figure \ref{fig:offline_latency} compares offline execution times of different RAG workflows on across different index types. While larger batches in offline scenarios are better suited to the modular design of existing frameworks, {\tool} still delivers significant performance gains, achieving speedups of 3.5$\times$ and 1.3$\times$ over LangChain and FlashRAG, respectively. The speedup is due to the efficient parallel execution across CPU and GPU further improve the hybrid system throughput.

\vspace{0.8em}

\noindent \textbf{Concurrency of different workflows}. We also evaluate {\tool}’s performance advantages under concurrent execution of requests from different RAG workflows (Figure~\ref{fig:hete_latency}). To simulate concurrent RAG workflows, we randomly interleave queries from the dataset and assign them to different workflows as input requests. We do not report results for LangChain, as all requests under concurrent execution exceed the SLO threshold. Concurrent workflows impose greater performance degradation on FlashRAG, particularly for complex workflows. In contrast, {\tool} maintains high efficiency under such concurrent workloads, achieving up to 5.5$\times$ latency reduction and 3.3$\times$ throughput improvement. This demonstrates {\tool}’s ability to seamlessly optimize performance across heterogeneous RAG workflows.


\noindent \textbf{Other LLMs}. We further evaluate \tool{} on larger LLMs, including Llama2-13B and OPT-30B. Figure~\ref{fig:other_model} shows that {\tool} achieves over 1.5$\times$ throughput improvement compared to existing frameworks. The gains are more pronounced under higher per-request latency (e.g., Llama2-13B), where fine-grained scheduling and pipelining more effectively mitigate inter-stage stalls. Larger models like Llama2-13B present distinct system-level behaviors. First, they introduce longer generation latencies and increased GPU memory pressure, which exacerbate pipeline imbalance between generation and retrieval stages. Second, different generation and reasoning capabilities across LLMs can lead to inconsistent behaviors for the same workflow and datasets. e.g. different number of iterations.

Despite these differences, {\tool} maintains consistent speedups across model scales, demonstrating its ability to adapt to varying model characteristics through dynamic stage partitioning and execution strategies. Besides, extremely large models paired with small databases tend to shift the performance bottleneck toward LLM inference—making retrieval coordination less impactful. We find that more balanced configurations yield better overall tradeoffs in both efficiency and accuracy. Importantly, our goal is not to optimize for large models per se, but to demonstrate that {\tool} generalizes across the diverse LLMs used in practice. In many RAG applications~\cite{lewis2020rag, borgeaud2022retro, gao2023ragsurvey}, small- and medium-scale models remain widely used and tend to benefit more from RAG, as external retrieval can effectively compensate for their limited parametric knowledge.



\begin{figure}
    \centering
    \includegraphics[width=\linewidth]{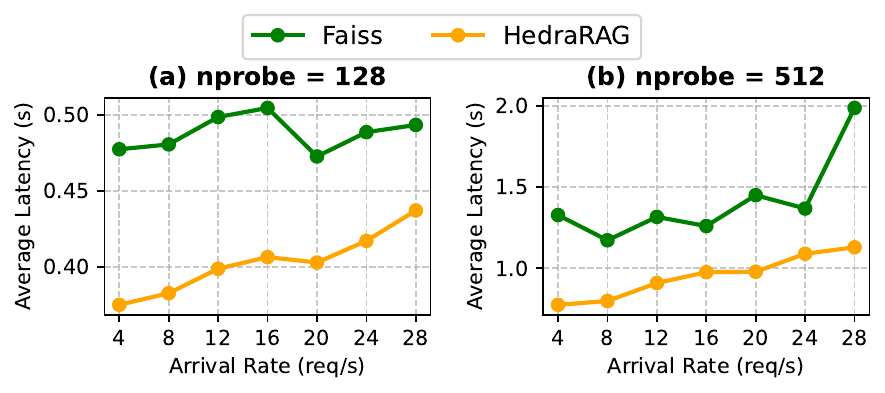}
    \caption{The fine-grained partitioning of {\tool} to improve vector database search latency.}
    \label{fig:ablation_faiss}
\end{figure}

\begin{figure}
    \centering
    \includegraphics[width=\linewidth]{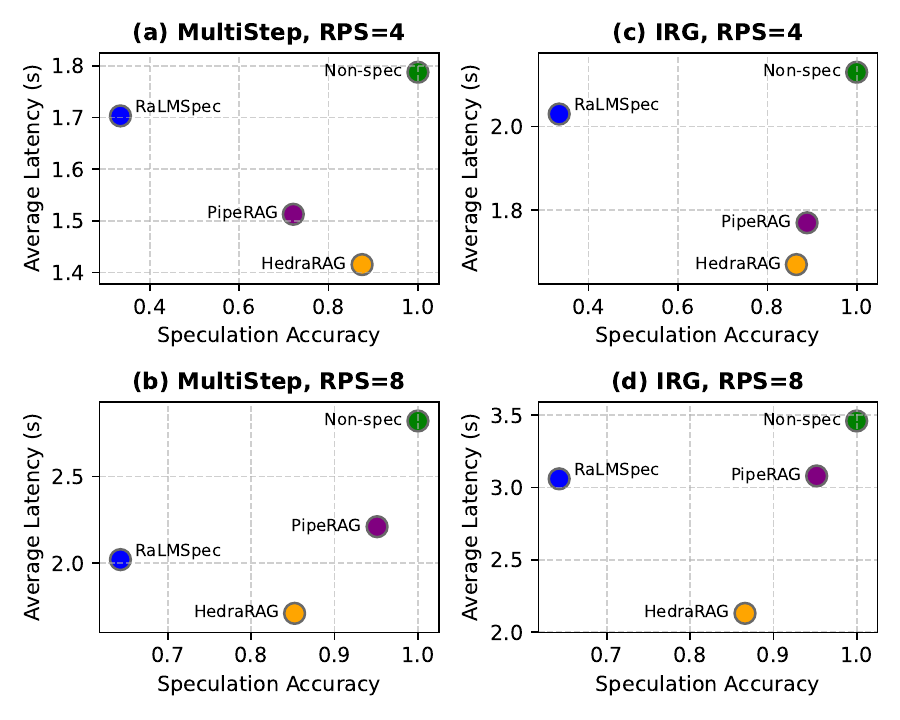}
    \caption{The speculation accuracy and latency comparison across existing speculation methods. RPS: request/second.}
    \label{fig:speculative_latency}
\end{figure}

\subsection{Performance Analysis}
\label{sec:technique effectiveness}

In this section, we perform an in-depth analysis and evaluation of the three optimization techniques of {\tool}.

\noindent \textbf{Dynamic partitioning and pipelining}. We evaluate how {\tool}’s sub-stage partitioning impacts vector database search latency in RAG serving scenarios. To simulate the fine-grained and non-batched search requests (typically come from the step-wise generation stages from different requests), we vary the request rate sent to the retrieval engine at the granularity of individual requests. As Figure\ref{fig:ablation_faiss} shows, {\tool} effectively improves search latency, achieving a reduction of 1.09$\times$ to 1.77$\times$. This improvement is due to the more fine-grained and dynamic batching, which minimizes the time for new requests to wait behind long-latency, coarse-grained search calls.



\begin{figure}[t]
    \centering
    \includegraphics[width=\linewidth]{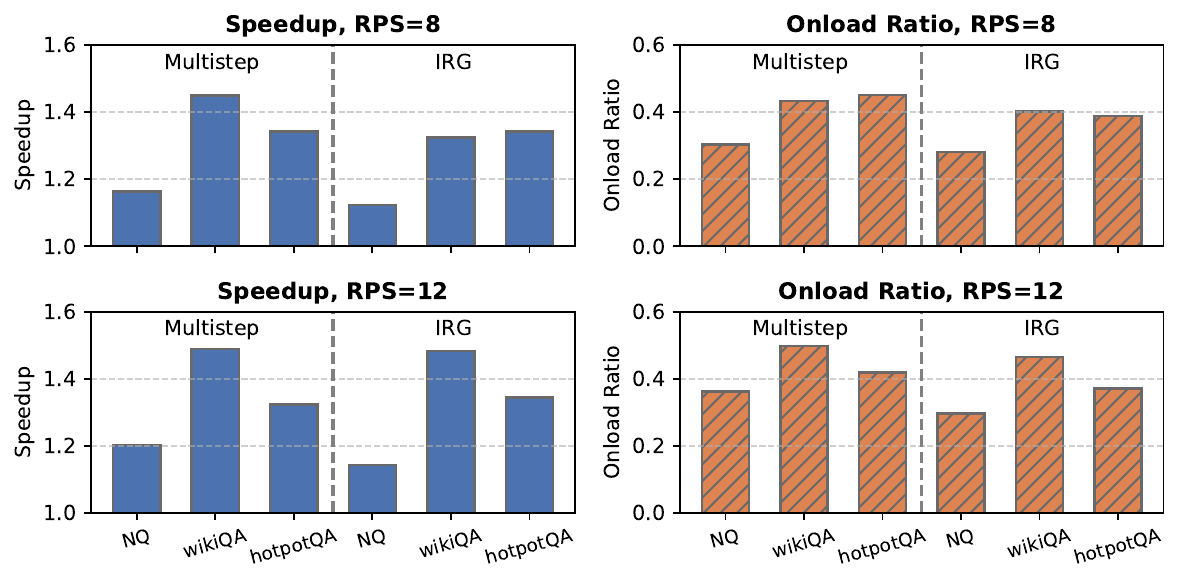}
    \caption{The speedups of GPU indexing and the hotspot cluster cache hit rate. $\mathit{nprobe}$ is set as 512.}
    \label{fig:gpu_indexing}
    \vspace{-0.8em}
\end{figure}

\noindent \textbf{Reordering and Speculation}. 
\label{sec:exp speculation} We compare {\tool}’s dynamic speculative execution strategy against existing approaches, including RaLMSpec and RAGCache, with respect to both speculation accuracy and end-to-end request latency.
Speculation accuracy is defined as the proportion of speculative generation steps in which the partially retrieved results match those produced by complete retrieval. As shown in Figure~\ref{fig:speculative_latency}, {\tool}'s similarity-aware reordering and dynamic speculation strategies yield a latency speedup ranging from 1.06$\times$ to 1.62$\times$ over prior methods. RaLMSpec, which relies solely on local cache contents, suffers from lower speculation accuracy and frequently incurs additional rollback overhead. PipeRAG adopts a more conservative speculation policy, as it does not leverage semantic similarity across retrieval stages, resulting in limited latency reduction. In contrast, {\tool} integrates similarity-aware reordering with a runtime-adaptive speculative execution mechanism that considers both RAG workflow heterogeneity and request-level workload dynamics.
This coordinated design leads to consistently higher speculation accuracy and more effective latency reduction across diverse workloads.


\noindent \textbf{Partial GPU Indexing}.
\label{sec:gpu indexing}
We evaluate the impact of GPU-based indexing on performance. The observed speedup is most pronounced when the retrieval stage incurs high CPU overhead, for example, when approaching system throughput limits (e.g., with $nprobe=512$ and request-per-second (RPS) between 8 and 12). Figure~\ref{fig:gpu_indexing} shows the GPU speedup as well as the probability that accessed clusters during retrieval are found in the GPU cache. GPU indexing yields a speedup ranging from 1.12$\times$ to 1.49$\times$, which correlates positively with the cache hit probability of accessed clusters. The hit rate varies across datasets. Intuitively, we attribute this variation to differences in topic skewness: compared to NQ, datasets like WikiQA and HotpotQA exhibit stronger access skewness, leading to higher cache reuse.

\section{Conclusion}
We present {\tool}, a co-designed LLM–vector search system for serving heterogeneous RAG workflows. \tool{} introduces {\graph}, a unified graph-based abstraction that expresses diverse workflow structures and enables generalizable optimizations. By defining transformation operations that model fine-grained sub-stage partitioning, semantic-aware speculative execution, and partial GPU index caching, \tool{} bridges the gap between high-level RAG heterogeneity and low-level LLM and vector search backends. Our evaluation shows that \tool{} consistently outperforms existing RAG systems across a range of models and workflows, demonstrating the value of rethinking system design to address the challenges of modern, heterogeneous AI pipelines.

\bibliographystyle{ACM-Reference-Format}
\bibliography{ref}

\end{document}